\documentclass[sigconf]{acmart}

\copyrightyear{2025}
\acmYear{2025}
\setcopyright{cc}
\setcctype{by}
\acmConference[UIST '25]{The 38th Annual ACM Symposium on User Interface Software and Technology}{September 28-October 1, 2025}{Busan, Republic of Korea}
\acmBooktitle{The 38th Annual ACM Symposium on User Interface Software and Technology (UIST '25), September 28-October 1, 2025, Busan, Republic of Korea}\acmDOI{10.1145/3746059.3747758}
\acmISBN{979-8-4007-2037-6/2025/09}

\usepackage{booktabs} 

\usepackage{exii-macros}

\usepackage{booktabs}
\usepackage[nameinlink]{cleveref}

\newcommand{\Name}{Visual story-writing\xspace}
\newcommand{\NaMe}{Visual Story-Writing\xspace}
\newcommand{\name}{visual story-writing\xspace}

\usepackage{acmart-taps}

\newcommand{\oposition}{\raisebox{-2.5pt}{\includegraphics{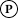}}}

\newcommand{\oassociate}
{\raisebox{-2.5pt}{\includegraphics{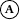}}}
\newcommand{\oconnect}{\raisebox{-2.5pt}{\includegraphics{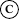}}}

\newcommand{\ounfold}{\raisebox{-2.5pt}{\includegraphics{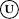}}}

\definecolor{modernblue}{HTML}{1A73E8}

\newcommand{\readtag}{\raisebox{-2.5pt}{\includegraphics{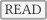}}}

\newcommand{\writetag}{\raisebox{-2.5pt}{\includegraphics{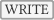}}}

\newcommand{\rwtag}{\raisebox{-2.5pt}{\includegraphics{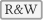}}}

\usepackage{fancyvrb}

\newenvironment{json}
  {\Verbatim[xleftmargin=2em]}
  {\endVerbatim \medskip}

\newenvironment{prompt}{
\par
  \noindent
  \ttfamily
  \setlength{\parindent}{0pt} 
  \setlength{\leftskip}{20pt}  

}{
  \par
  \medskip
} 

\begin{document}

\tolerance=400 

%
\title[\NaMe]{\NaMe: Writing by Manipulating\\Visual Representations of Stories}

\author{Damien Masson}
\orcid{0000-0002-9482-8639}
\affiliation{%
  \institution{Université de Montréal}
  \city{Montreal}
  \state{Quebec}
  \country{Canada}
}
\email{damien.masson@umontreal.ca}

\author{Zixin Zhao}

\orcid{0000-0002-8636-1987}
\affiliation{%
  \institution{University of Toronto}
  \city{Toronto}
  \state{Ontario}
  \country{Canada}
}
\email{zzhao1@cs.toronto.edu}

\author{Fanny Chevalier}
\orcid{0000-0002-5585-7971}
\affiliation{%
  \institution{University of Toronto}
  \city{Toronto}
  \state{Ontario}
  \country{Canada}
}
\email{fanny@dgp.toronto.edu}



\begin{abstract}
We define ``\name'' as using visual representations of story elements to support writing and revising narrative texts. To demonstrate this approach, we developed a text editor that automatically visualizes a graph of entity interactions, movement between locations, and a timeline of story events. Interacting with these visualizations results in suggested text edits: for example, connecting two characters in the graph creates an interaction between them, moving an entity updates their described location, and rearranging events on the timeline reorganizes the narrative sequence. Through two user studies on narrative text editing and writing, we found that visuals supported participants in planning high-level revisions, tracking story elements, and exploring story variations in ways that encourage creativity. Broadly, our work lays the foundation for writing support, not just through words, but also visuals.
\end{abstract}

%
%



\maketitle



\section{Introduction}
Creative writers manage many moving parts, from character arcs and causal chains to spatial coherence and narrative timing~\cite{balNarratologyIntroductionTheory1997, nortonDevelopmentalEditingHandbook2011}. Keeping track of everything is challenging, and it gets even more complicated when experimenting with story ideas~\cite{zhao2025making}. Modifications require careful planning and multiple rounds of edits.

\begin{figure}[t]
    \includegraphics[width=\columnwidth]{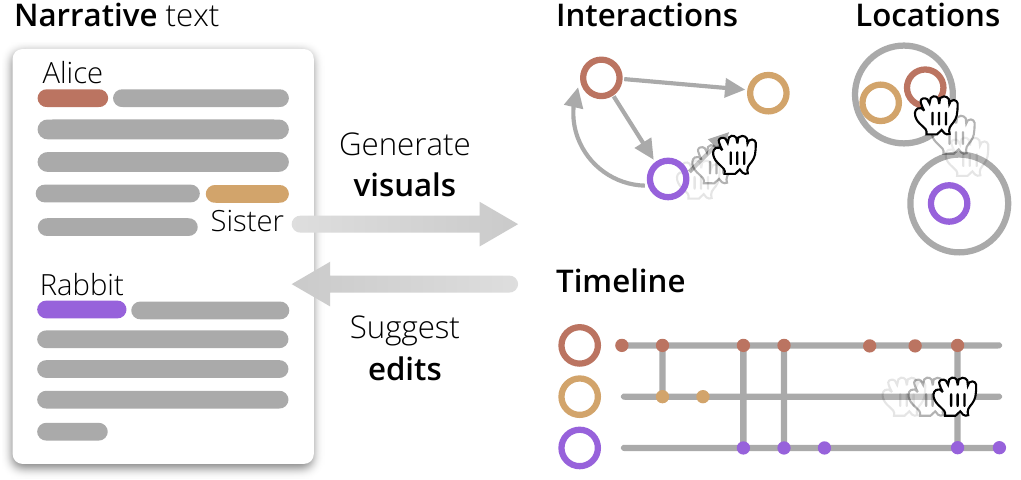}
    \caption{\Name supports writing by generating visualizations that help review the story and that can be manipulated to suggest edits to the narrative text.}
    \label{fig:abstract}
\end{figure}

Consider a task as simple as changing the location of a character in a story: the cat no longer goes to the barn but instead wanders about the lake.
Simply replacing all instances of \textit{``barn''} would cause inconsistencies unless descriptions and actions pertaining to the barn are also updated to reflect the new context of the lake. Moreover, the fact that the cat wanders about the lake does not mean that the barn needs to be removed entirely from the story. 
Instead, before revising the text, writers must identify references to the barn and tediously review the other characters' locations to maintain spatial and semantic consistency.
This busywork is prone to errors, and writers often resort to creating external documents, such as maps, spreadsheets, and timelines, that they painstakingly update to keep track of characters, events, and locations~\cite{zhao2025making, ackermanEfficientWriterUsing2017, neuwirthRoleExternalRepresentation1989, SaveTheCat, Articy}. Large language models (LLMs) could help, but expressing specific intents with a prompt is difficult~\cite{massonDirectGPTDirectManipulation2024, zamfirescu-pereiraWhyJohnnyCant2023}, and trying to explain which barn and which cat to edit will leave much room for misinterpretation. Further, the suitability of the changes made by the generative model could be difficult to verify.

This challenge stems from the mismatch between the reasoning and representation spaces: in our \textit{barn} example, the writer does spatial reasoning with a textual representation. A better approach would be to use a spatial representation, such as a map of the story world, where changing the location of a character becomes as simple as moving it on the map. 
This same argument applies to all other aspects of a story, such as time, events, and characters, for which we posit superior representations can be designed.

Therefore, we propose using representations that match the reasoning space. 
We define \textbf{\name} as the use of visual representations of story elements as both a reviewing tool and an input medium for expressing writing intents (\Cref{fig:abstract}).
\rev{This complements the writing workflow of writers with various experience levels. For example, an experienced writer may prefer editing the story's text, in which case the visual representations updates, offering a visualization to help the writer track story elements. Alternatively, a beginner writer could manipulate the visual representations through direct manipulation to explore story variations.}

We articulate a framework grounded in narratology to design new visual representations of stories (\S\ref{sec:taxonomy}).
Applying our framework, we explore the potential of \name through a prototype software (\S\ref{sec:tool}) with three illustrative visual representations: (1) a diagrammatic view of the entities in the story and their relationships, allowing to add, remove, and edit characters and add new actions between them; (2) a spatial view of the entities in the story, allowing to add new locations and move entities; and (3) a timeline of the events in the story, allowing to reorder the narrative, quickly find specific scenes, and precisely modify the entities and locations of the selected scenes by leveraging the other views.

We validate this approach through two user studies looking at the different components of the writing process~\cite{flowerCognitiveProcessTheory1981}. Results from our first study revealed that visuals helped plan high-level revisions, search, and promote critical reflection over having only text. 
Results from our second study show the potential of \name to express editing intents and help writers explore story variations in a way that supports creative expression.
This points to the potential of \name and encourages further exploration.

\section{Related Work}
\Name draws inspiration from work on visualizing stories and efforts to give alternative representations to edit content.

\subsection{Visualizing Stories} 
Visual representations of stories and story patterns have a long-standing presence in writing, with staple examples including Freytag's pyramid story~\cite{freytagTechniqueDramaExposition2004} and abstract diagrams capturing the ``shape of the story''~\cite{popeGoodScriptsBad1998, shape-story-vonnegut, reaganEmotionalArcsStories2016}. Such visual representations are not only useful for teaching the craft of storytelling but are also considered useful aids to support the writing process. Referring to the visual representations, screenplay writer E. Williams says: \emph{``I find story shape to be a great tool [...] to make sure my story holds together''}~\cite{williamsHowViewAppreciate2018a}.

There exist many techniques to visualize stories, spanning from classical charts, like word clouds, to more artistic visualizations like calligrams~\cite{brathSurveyingWonderlandMany2021}. \rev{Often, these visualizations attempt to highlight a specific facet of the story (e.g., emotions~\cite{maharjanLettingEmotionsFlow2018}, 
motion~\cite{chungToytellerAIpoweredVisual2025},
fortune~\cite{shape-story-vonnegut, chungTaleBrushSketchingStories2022}, or topic recurrence~\cite{zhuPersistentHomologyIntroduction2013}). For example, Storyline visualizations~\cite{munroeMovieNarrativeCharts2009, tanahashiDesignConsiderationsOptimizing2012} depict the evolution of the relationships between characters through timelines.} StoryCake~\cite{qiangStoryCakeHierarchicalPlot2016} uses a hierarchical plot to visualize the structure of nonlinear stories. Story Curves~\cite{kimVisualizingNonlinearNarratives2018} visualizes non-linear narratives by showing both the order in which the events are told and the story's chronological order.
StoryPrint~\cite{watsonStoryPrintInteractiveVisualization2019} shows the scenes and character emotions through circular timelines. Each of these visual representations presents the ability to support readers, writers, and analysts alike in appreciating or even analyzing the stories from a specific lens.

Other story visualizations have been used in tools to help with script and story writing. For example, CARDINAL~\cite{martiCARDINALComputerAssisted2018} helps screenwriters visualize their script through different views, including a visualization of the interactions between characters and animated 2D and 3D views to playout the scene and replay character movements. Similarly, Portrayal~\cite{hoquePortrayalLeveragingNLP2023} attempts to help writers develop their characters by extracting and presenting information related to the characters in interactive visualizations.
\rev{Our work seeks to inform the design of story visualizations by proposing a new framework. 
This framework helps generate representations that might serve as a medium to edit the story, a topic we discuss next.}

\subsection{Editing Using Alternative Representations}
Visualizations are increasingly used to not only encode data but also collect and modify new data~\cite{bressa2024-input-vis}. In our work, visual representations can be edited to modify the narrative text in addition to exploring and analyzing it, making it an input visualization.

One domain which has embraced the use of input visualizations is computer programming. For example, visual programming~\cite{myersVisualProgrammingProgramming1986, burnettVisualProgramming1995} allows generating programs by manipulating elements in more than one dimension, code projections~\cite{gobertLorgnetteCreatingMalleable2023, koBaristaImplementationFramework2006} augments programming languages with tools and widgets that are alternative representations of fragments of code, and bi-directional programming~\cite{mayerBidirectionalEvaluationDirect2018a, hempelSketchnSketchOutputDirectedProgramming2019} allows manipulating the output of a program to edit the program itself. Similar to our work, the motivation behind these approaches is that some tasks are easier done using different representations: it is easier to find a colour using a colour picker than by typing its hexadecimal code, and it is easier to move an object through direct manipulation~\cite{shneidermanDirectManipulationStep1983} than by typing coordinates.

In the creative writing domain, the idea of manipulating projections or visual representations to edit stories is largely untapped despite the reliance on visual representations to capture story concepts~\cite{freytagTechniqueDramaExposition2004, shape-story-vonnegut,zhao2025making}. Instead, most writing support tools ask users to manipulate narratives using text, typically through prompts or suggestions~\cite{calderwoodHowNovelistsUse2020, buschekImpactMultipleParallel2021, yuanWordcraftStoryWriting2022, dangChoiceControlHow2023, leeDesignSpaceIntelligent2024, rezaABScribeRapidExploration2024}. Some exceptions include the proposal of \citet{kempenAuthorEnvironmentsFifth1987} to manipulate the syntax tree of a sentence to restructure it or the proposal to directly drag the words to reorder them~\cite{arnoldGenerativeModelsCan2021}. Beyond syntactic level modifications, \citet{dangTextGenerationSupporting2022} proposed a system that allows reorganizing, merging, and removing paragraphs by manipulating textual summaries. 
\rev{Moving even further from text as an input medium, the machine learning task of ``visual storytelling'' consists in generating a story from a sequence of images~\cite{huangVisualStorytelling2016, hsuVisualStoryPostEditing2019, wangShowRewardTell2018, hsuHowUsersEdit2019, bensaidFairyTailorMultimodalGenerative2021}.} Similarly, TaleBrush~\cite{chungTaleBrushSketchingStories2022} leverages a canvas where sketching allows modulating the fortune of characters in a story, whereas Textoshop~\cite{masson2025Textoshop} offers features inspired by drawing software such as tone pickers and text layers.

Closest to our approach are systems that provide an editable alternative view of the text. For example, VISAR~\cite{zhangVISARHumanAIArgumentative2023} generates a visual outline of an argumentative text, which can be edited to change the structure of the argument. \rev{Toyteller~\cite{chungToytellerAIpoweredVisual2025} generates story text based on character motions, as if the characters were toys.} Similarly, XCreation~\cite{yanXCreationGraphbasedCrossmodal2023} supports creativity and cross-modality edits through the use of clip-arts and a relationship diagram as an intermediate representation between text and images.

Our work contributes to this line of research by formalizing \emph{\name}. The concept is rooted in the same principles as visual programming and code projections, positing that elements of a story become easier to understand and manipulate when presented in alternative forms. Specifically, we propose a framework grounded in structural narratology theories to inform the design of visualizations and extend the concept to all story elements, including events, like XCreation~\cite{yanXCreationGraphbasedCrossmodal2023}, but also space and time.

\section{A Framework for Story Constructs}
\label{sec:taxonomy}
To inform the design of visual representations that support the \name workflow, we articulate a generative framework.
Our framework builds upon narrative theories by structuralists like G\'{e}rard Genette~\cite{genette1990fictional} and Mieke Bal~\cite{balNarratologyIntroductionTheory1997, balNarratologyPractice2021}. Elements of stories outlined by structuralists~\cite{balNarratologyIntroductionTheory1997, balNarratologyPractice2021} match what we observed in existing story visualizations, such as organizing story elements based on time~\cite{munroeMovieNarrativeCharts2009, watsonStoryPrintInteractiveVisualization2019}, space~\cite{martiCARDINALComputerAssisted2018}, or a combination thereof~\cite{hulsteinGeoStorylinesIntegratingMaps2023}. Story elements do not work in isolation. Rather, many important aspects of stories stem from the intricate weaving of these elements, such as the movement of characters across locations over time. 

We propose an \emph{operator} mechanism to combine elements into such meaningful composites. 
Our operators are applied sequentially to \textit{story elements} to form a story construct that can then be represented and manipulated. This framework can describe story constructs in existing visualizations and also generate new ones.

\begin{figure*}
\includegraphics[width=\textwidth]{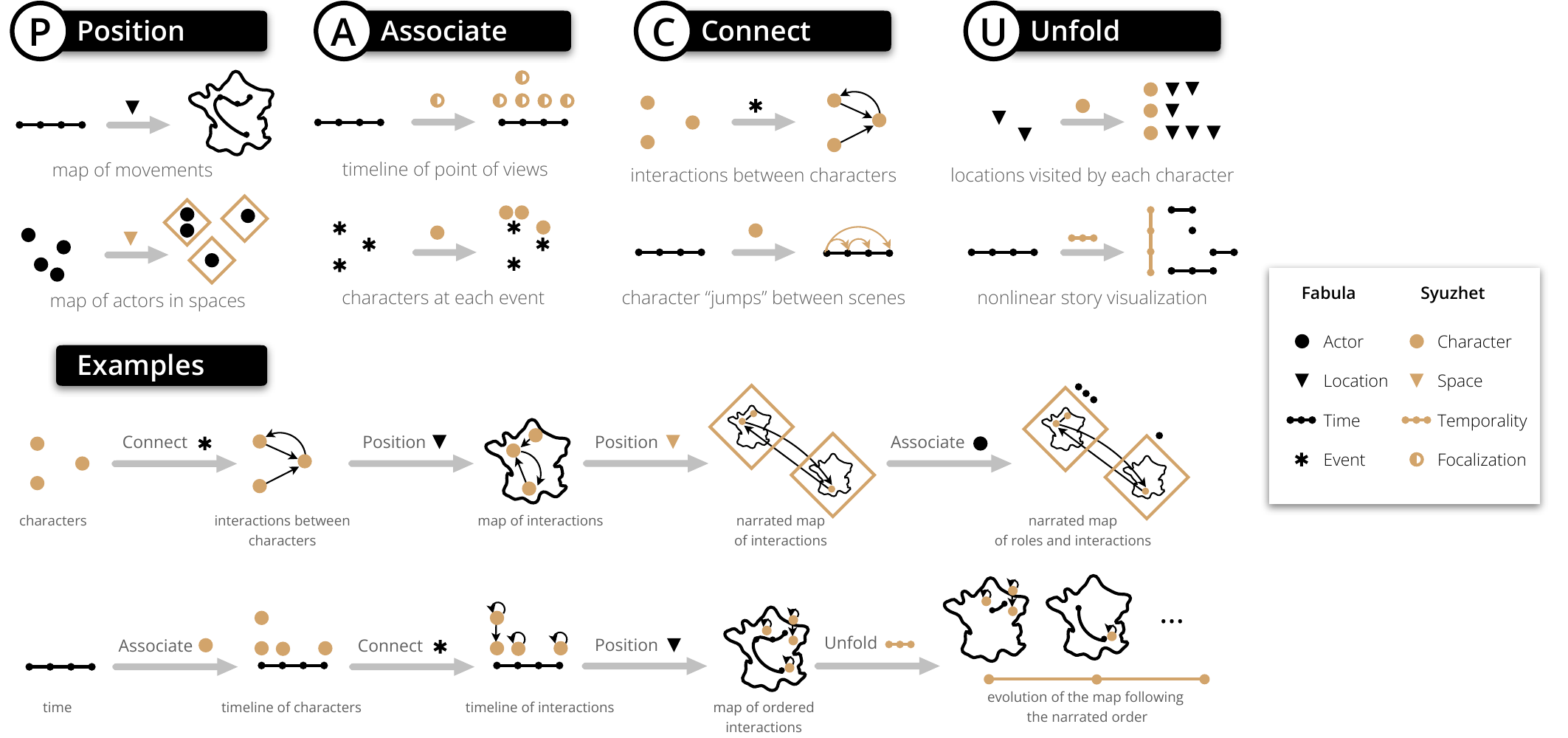}
   \caption{We propose a framework grounded in narratology to generate story constructs that can support \name. First, select an element from the fabula or syuzhet and then apply operators sequentially: \oposition{} \textsc{position} to place elements spatially; \oassociate{} \textsc{associate} to add new elements; \oconnect{} \textsc{connect} to add edges; and \ounfold{} \textsc{unfold} to duplicate and organize elements.}
   \label{fig:operator}
\end{figure*}

\subsection{Story Elements}
Story elements are the fundamental building blocks of story constructs found in story visualizations. 
Within narratology, the early Russian formalists separated the chronological order of the story (\textit{fabula}) from the order of the plot or how it is represented to readers (\textit{syuzhet})~\cite{liveleyNarratology2021, balNarratologyIntroductionTheory1997}. Structuralists like Genette and Bal borrowed these concepts to analyze narrative structurally. Bal~\cite{balNarratologyIntroductionTheory1997,balNarratologyPractice2021} describes a fabula, the factual events within a story, to consist of four main elements: actors\footnote{Actors refer to ``agents that perform actions. They are not necessarily human''~\cite{balNarratologyIntroductionTheory1997}. To avoid confusion, we use the term \textit{entity} instead in our prototype system.}, time, location, and events. \rev{The syuzhet, or how the story is told, also consists of four elements: characters, temporality, space, and focalization. See \cref{tbl:storyelements} for examples.} 

\begin{table}[h]
\caption{\rev{The eight story elements of our framework, extracted from fabula and syuzhet elements described by Bal~\cite{balNarratologyIntroductionTheory1997,balNarratologyPractice2021}}}

\begin{tabular}{ll}
\toprule
Element of Fabula & Element of Syuzhet  \\
\midrule
Actor (e.g., hero, villain) & Character (e.g., Alice) \\
Location (e.g., Alice's house) & Space (e.g., home, eerie) \\
Time (chronological) & Temporality (narrated order) \\
Event (what happens) & Focalization (point of view) \\
\bottomrule
\end{tabular}
\label{tbl:storyelements}
\end{table}

Although elements of fabula and syuzhet may seem similar, they refer to distinct concepts. For instance, characters are concrete entities (e.g., Alice, the white rabbit), whereas actors are abstract roles or functions characters can take (e.g., hero, villain). Time corresponds to the chronological timeline of events, whereas temporality corresponds to how those events are narrated (e.g., through flashbacks, ellipsis). And locations are concrete, physical places (e.g., Alice's house, a forest), whereas spaces refer to the setting and the narrated representation of these locations (e.g., home, eerie). We describe \name as the use of visuals that represent and manipulate story constructs, i.e. composites of these eight elements.

\subsection{Operators}
Our operators are functions that combine story elements (operands) into higher-level story constructs. They can be chained to create more elaborate constructs. Based on existing story visualizations, we derived four binary operators acting on operands $x$ and $y$\footnote{Note that our operators are noncommutative. Thus, while $y$ is always one of the eight story elements, $x$ can be a composite of elements resulting from a chain of operations.}:

\begin{itemize}
    \item \textsc{position} ($x$ \oposition{} $y$) places elements of $x$ based on the story element $y$ (\Cref{fig:operator}.\oposition{}). For example, positioning \textit{time} by \textit{locations} results in a map of the movements in the story. Note that the only valid $y$ operands are \textit{location} and \textit{space}. 
    \item \textsc{associate} ($x$ \oassociate{} $y$) adds the elements of $y$ and associates them to the elements of $x$ (\Cref{fig:operator}.\oassociate{}).
    For example, associating \textit{time} with \textit{focalization} creates a timeline of point of view.
    \item \textsc{connect} ($x$ \oconnect{} $y$) adds edges between elements of $x$, according to that of $y$  (\Cref{fig:operator}.\oconnect{}). For example, connecting \textit{characters} by \textit{events} creates a graph of character interactions.
    \item \textsc{unfold} ($x$ \ounfold{} $y$) duplicates and organizes elements of $x$ based on that of $y$  (\Cref{fig:operator}.\ounfold{}). This essentially duplicates the $x$ elements according to each element of $y$. For example, unfolding \textit{locations} by \textit{characters} creates a list of locations visited by each character.
\end{itemize}

\subsection{Describing Existing Story Visualizations} \label{sec:framework-demo}
\rev{We test our framework by describing existing story visualizations.

\begin{itemize}
\item \textit{Storyline }~\cite{munroeMovieNarrativeCharts2009} shows character interactions: start with \textit{time}, \textsc{unfold} by \textit{characters}, and \textsc{connect} by \textit{events}.

\item \textit{StoryCurve}~\cite{kimVisualizingNonlinearNarratives2018} shows nonlinear narratives: start with \textit{time}, \textsc{unfold} by \textit{temporality} to visualize the non-linear narrative, \textsc{associate} with \textit{locations} (using colours) and \textsc{associate} with \textit{events}. 

\item \textit{StoryPrint} ~\cite{watsonStoryPrintInteractiveVisualization2019} shows scenes, character presence and emotions: start with \textit{time}, \textsc{unfold} by \textit{characters}, and \textsc{associate} with \textit{events} or other elements (StoryPrint also encodes emotions, which goes beyond the scope of structural narratology). 

\item \textit{Geo-Storylines}~\cite{hulsteinGeoStorylinesIntegratingMaps2023} shows locations of characters over time: start with Storyline, \textsc{position} by \textit{locations}. The map glyph variation: start with \textit{locations}, \textsc{position} by \textit{location}, \textsc{unfold} by \textit{time}, and \textsc{associate} with \textit{characters}. 

\item \textit{``All fights from Dragon Ball Z''}\footnote{A non-academic example}~\cite{cinnamonAllFightsDragon2017} shows the battles between characters within a saga: start with \textit{events} (only the fights), \textsc{unfold} by \textit{time}, and \textsc{connect} by \textit{characters}.
\end{itemize}}

\subsection{Generating New Story Constructs}\label{sec:framework_new_viz}
\Cref{fig:operator} shows examples of chaining operators to create more complex story constructs. The number of unique combinations of operators and elements is vast. Our goal is not to argue which specific story construct is optimal for writers nor which visual design is best to represent these constructs---we leave it to future work to explore variations within these categories (and beyond). Instead, our work focuses on the potential of \name, i.e. the use of alternative representations of stories, as an input medium for expressing intent. As such, we designed three simple visual representations of story constructs interconnected through interactivity~\cite{northTaxonomyMultipleWindow2001, wang_baldonado_guidelines_2000}: a timeline of entities and events (start with \textit{time}, \textsc{unfold} by \textit{characters} and \textsc{connect} by \textit{events}); a graph of actions and entities (start with \textit{characters} and \textsc{connect} by \textit{events}). And a view of the locations (start with \textit{characters} and \textsc{position} by \textit{locations}). 
\rev{
There are other important components to a story, such as character traits and emotions, but in Bal's narratology model~\cite{ balNarratologyIntroductionTheory1997, balNarratologyPractice2021}, these are properties of the story elements. One way to represent them within the framework is as an attribute of a story element.
}

\rev{
\subsection{Recommending Interactions and Visuals}
The framework can be further expanded to prescribe interactions and visual representations. For example, applying the \textsc{position} operator suggests that the placed elements can be dragged to express a change of location; the \textsc{associate} operator suggests a detail-on-demand type of interaction to view the associated elements; the \textsc{connect} operator suggests an interaction to remove and create links between elements; and the \textsc{unfold} operator suggests elements can be dragged to be reordered or reassigned. Similarly, the framework could also prescribe visual representations by mapping operations to specific visual channels.
}

\section{The \NaMe Prototype}
\label{sec:tool}
\begin{figure*}[t]
\includegraphics[width=\textwidth]{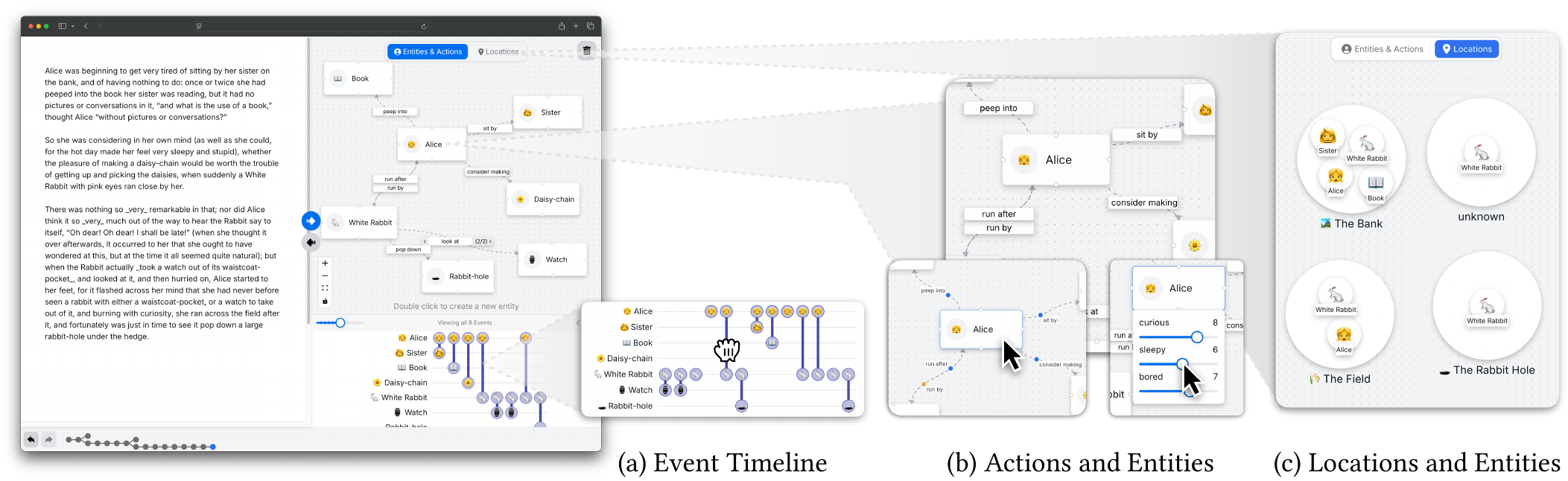}
   \caption{The \name system used on the first three paragraphs of \textit{Alice's Adventures in Wonderland}: (a) the event timeline allows reordering the events; (b) the actions and entities view allows editing the characters' traits and adding or removing entities and actions; (c) the locations and entities view allows moving entities and creating new locations}
   \label{fig:teaser}
\end{figure*}

To test the potential of \name, we developed a prototype system, shown in \Cref{fig:teaser}, with three visual representations informed by our framework: a diagrammatic view of the entities and actions, a spatial view of the locations and entities, and a timeline of the events (\cref{sec:framework_new_viz}). The focus of the prototype was to help writers explore variations and perform spatial or temporal edits via \name. As such, the system allowed specifying edits through direct manipulations of the graphical elements~\cite{shneidermanFutureInteractiveSystems1982, hutchinsDirectManipulationInterfaces1985} and followed the design principles of creativity support tools to support exploration~\cite{shneidermanCreativitySupportTools2006}. Below, we detail the system design and features \rev{(marked with \readtag{} when the feature helps review the story, \writetag{} when it helps edit the story, and \rwtag{} when it helps for both editing and reviewing)}. All screenshots were taken using the system on the first three paragraphs of the story \textit{Alice's Adventures in Wonderland} by Lewis Carroll.

\subsection{Entities and Actions View}\label{sec:entityView}
This representation shows all the entities in the story as nodes with their name and an emoji representing them (\Cref{fig:teaser}.b). The actions between the entities are represented as directed edges with a label describing the action in one or two words. By default, the nodes are arranged using a force layout to prevent overlaps, but users can rearrange the position of the entities by dragging them.

\begin{figure*}[h]
    \includegraphics[width=\textwidth]{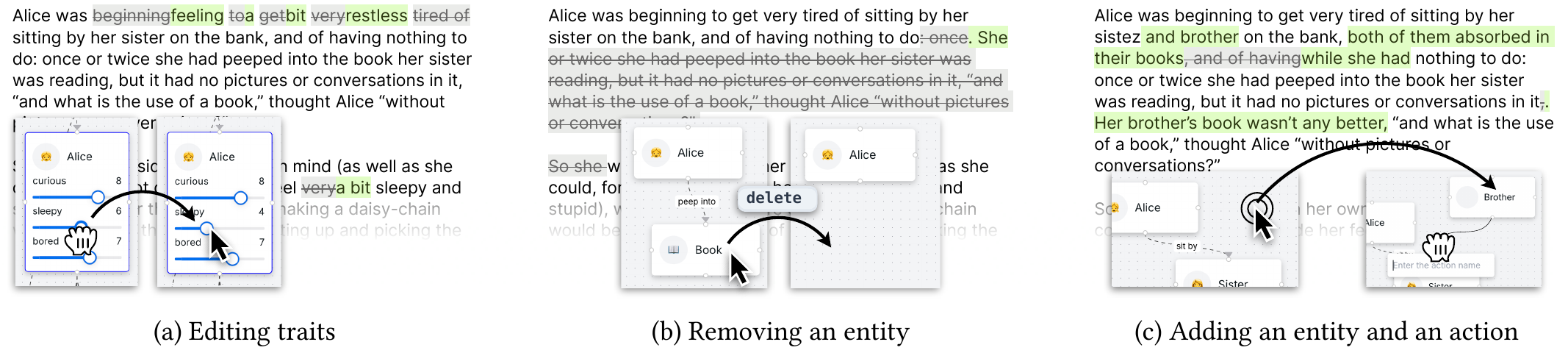}
    \caption{The entity view allows modifying entities: (a) selecting an entity opens up character traits that can be modified to edit the story; (b) an entity can be removed in which case the story is edited so that the entity and its actions are removed; and (c) a new entity can be created and connected to other entities to create actions}
    \label{fig:entityView}
\end{figure*}

\subsubsection{\rwtag{} Editing the Traits of an Entity} Selecting an entity opens a menu with traits similar to a list of personality traits which would be present on a character sheet (e.g., ``curious'', ``adventurous'', ``chatty''). The intensity of each trait is rated on a scale of 1 to 10. Moving the slider beneath the trait results in editing the text of the story to reflect the new intensity of the trait (\Cref{fig:entityView}.a).

\subsubsection{\writetag{} Adding and Removing an Entity} Double-clicking on the canvas opens a text input box, which, once filled, creates an entity with the given name at the location of the mouse pointer.
Selecting an entity and then pressing ``delete'' or ``backspace'' on the keyboard removes the entity. This results in updating the text of the story so that the entity is effectively removed  (\Cref{fig:entityView}.b).

\subsubsection{\writetag{} Adding and Removing Actions between Entities} Creating an edge between two nodes opens a text input field to enter a new action. The direction of the edge indicates the source and target of the action. Connecting a node to itself means the action is for the same character. This helps represent actions done by an entity with no target (e.g., walking or thinking). 
Double-clicking the label of an action edits it. Similarly, selecting an action and pressing ``delete'' or ``backspace'' removes it.
As soon as the edge is created, edited, or deleted, the story text is updated to reflect the change  (\Cref{fig:entityView}.c).

\subsubsection{\readtag{} Overlapping Edges and Animations} To declutter the view, overlapping edges are grouped so that only the first action is shown alongside a counter. Users can cycle through the events by clicking the next and previous buttons. Alternatively, hovering over an entity shows animated dots coming and going from the entity, with the name of the action beside them.

\subsection{Locations and Entities View}
The location view is a spatial representation accessed through a tab (\Cref{fig:teaser}.c). It displays all the locations in the story as nodes with their name and an emoji representing them. Similar to the entities and actions view (\Cref{sec:entityView}), the view shows all entities as smaller nodes, but their position depends on their order of appearance in the story. Entities that go to different locations in the story are duplicated to be represented in all their locations.

\begin{figure*}[h]
    \includegraphics[width=\textwidth]{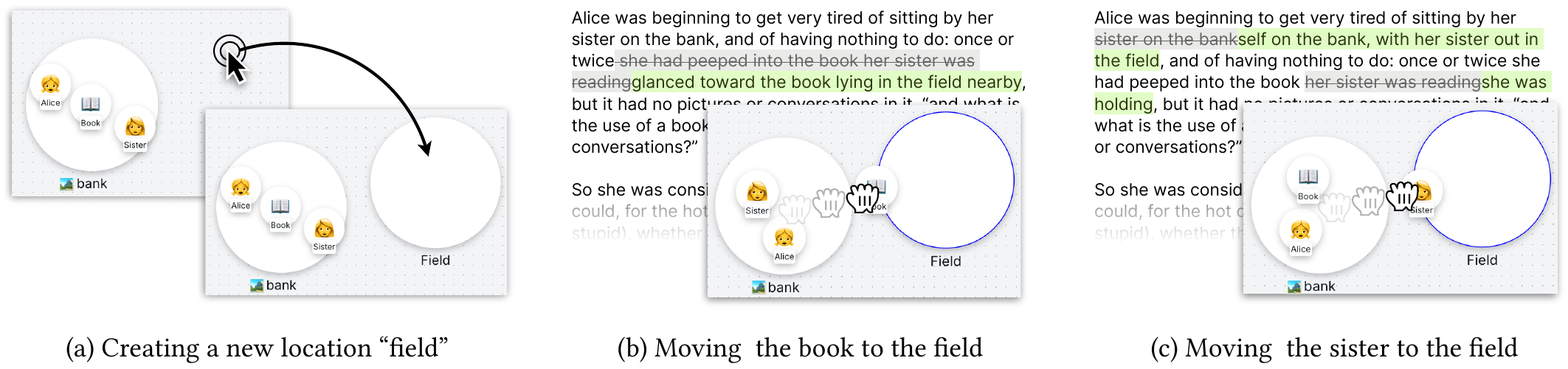}
    \caption{The location view allows moving entities: (a) a new location ``field'' is created through a double click; (b) the entity ``book'' is moved to the field through a drag-and-drop; and (c) the sister is moved to the field instead}
    \label{fig:locationView}
\end{figure*}
\subsubsection{\writetag{} Adding a Location} Double-clicking anywhere on the canvas opens a text input field to create a new location (\Cref{fig:locationView}.a). Once the location is created, entities can be moved to it.

\subsubsection{\writetag{} Moving an Entity} Entities can be dragged and moved around the view. If the entity is released on top of a location, then that entity is moved to that location (\Cref{fig:locationView}.b-c). Otherwise, the entity goes back to its original location. Once the entity is moved, the story text is updated to reflect the change in location.

\subsection{Timeline of Events View}
The timeline view is a temporal representation showing the events in the story represented as vertical lines with the emojis of the entities involved in the event on either side of the line (\Cref{fig:teaser}.a). These lines are organized one after the other based on the order in which they are presented in the text (\textit{temporality}). The timeline view is always visible. It allows selecting one or multiple events, in which case the other views are updated to fade out the entities and locations not involved in the selected events. Selecting events also forces subsequent modifications to impact only the selected events.

\begin{figure*}[h]
    \includegraphics[width=\textwidth]{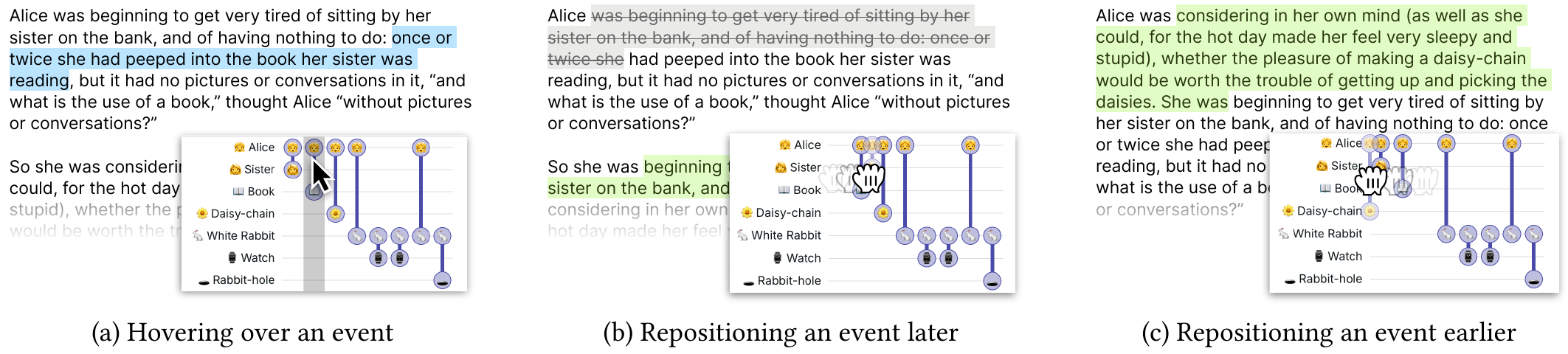}
    \caption{The timeline view allows finding and reordering the events in the story: (a) hovering over an event highlights the corresponding text; (b) the event about Alice sitting beside her sister is moved later in the story; (c) the event about Alice considering making daisy-chains is moved earlier in the story}
    \label{fig:timelineView}
\end{figure*}

\subsubsection{\rwtag{} Selecting Events} The timeline also allows finding specific scenes and selecting them to allow precise editing. Hovering over events highlights the corresponding sentence in the story (\Cref{fig:timelineView}.a) and the corresponding entities and locations in the event and location views. Clicking selects the event, and dragging selects multiple events. Once events are selected, future visual edits will target these selected events. For example, selecting an event and then moving an entity in the location view updates the story text so that the entity moved exactly during the selected event.

\subsubsection{\writetag{} Reordering Events} Dragging an event slides it horizontally (\Cref{fig:timelineView}.b-c). Once moved, the story text updates so that the event happens when indicated in the timeline.

\subsection{Bi-directional Editor}
The rest of the interface consists of a text editor on the left, a history tree at the bottom, and interface buttons at the centre to refresh the visuals or rewrite the story (\Cref{fig:teaser}).

\subsubsection{\readtag{} Highlighting Visual Elements and Sentences on Hover}
Similar to how hovering over events in the timeline highlights the corresponding sentences, hovering over the sentences in the text editor highlights the corresponding events, entities, and locations in the different views. This helps find scenes. Similarly, placing the caret in one sentence will act as if the events in that sentence were selected in the timeline. As such, subsequent edits to the visual representations will be restricted to that sentence.

\subsubsection{\readtag{} Updating the Visual Representations from the Text}
When the text is updated manually, the visual representations might become desynchronized, showing the story as it was before the modification. In this case, the refresh button becomes blue to indicate that the visual model might be outdated. Clicking it re-extracts the information from the text and refreshes the views.

\subsubsection{\writetag{} Writing the Story from the Visual Representations} By clicking the ``refresh from visuals'' button (arrows between the two views in \Cref{fig:teaser}), the story is regenerated from scratch using only the visual representations as reference. The regenerated story preserves the same events, locations, and entities, but everything else is rewritten. This allows for specifying the skeleton of a story and quickly exploring different ways of phrasing it.

\subsubsection{\writetag{} Track changes}
Changes made by editing the visual representations are tracked, such that removed text is struck while additions are highlighted in green. This allows writers to locate and review the changes before accepting (or rejecting) them.

\subsubsection{\writetag{} History Tree} 
To make edits easily reversible and facilitate the exploration of alternatives, the system implements a history tree to store different versions of the story and return to them at any time. When a modification happens, be it by editing the visual representation or editing the text, a snapshot of the story and visual representations is added to the tree. When the writer selects a previous snapshot, the story is reverted to that snapshot. If, after selecting a previous snapshot, a modification is made, then a new branch is created. This allows writers to navigate between branches and versions by clicking on the nodes of the history tree.

\subsection{Implementation}
We implemented the \name system using TypeScript and React~\cite{ReactJavaScriptLibrary2013} for the interface, NextUI~\cite{NextUI} for the graphical components, Slate.js~\cite{SlateJS} for the text editor, and the OpenAI library for prompting large language models~\cite{openaiOpenAINodeAPI2023}. Modifications to the visual representations are turned into engineered prompts sent to OpenAI's ``GPT4-o'' model. A live demo and the source code are accessible online: {\color{modernblue}\url{https://github.com/m-damien/VisualStoryWriting}}

\subsubsection{Extracting Entities, Locations, and Events} In our tests, the na\"{i}ve approach of asking an LLM to extract the necessary information results in slow and incomplete extractions. Instead, we first ask the model to extract the entities and locations in the story. We then split the text into sentences and parallelize requests by giving the model the whole story but asking it to extract only the information within each sentence (\cref{appendix:extraction}). As such, a text with 30 sentences sends 30 smaller prompts in parallel. This leads to faster and more exhaustive extractions. It also helps associate sentences with events. All requests use a structured JSON output.

\subsubsection{Engineered Prompts to Edit the Story} The system relies on engineered prompts such as splitting requests to make them faster, giving the model only a subset of the text to force localized edits, and giving the model the previous abstract representation of the story and the new one to help it understand the edits. \rev{See the appendix and/or the source code for more details.}

\section{Study 1: Planning Using the Visualizations}
\label{sec:study1}
In our initial user study, we assessed how story visualizations aid the planning and reviewing phases of writing~\cite{flowerCognitiveProcessTheory1981}. \rev{To ensure the study focused solely on visualization, the prototype only supported the \readtag{} features.} We then compared this read-only version to a baseline text-only interface, asking participants to answer high-level planning and reviewing questions. \rev{Because no text had to be produced, we recruited participants with various creative writing experiences to reflect the many potential target users of the tool}

\subsection{Method}

\subsubsection{Participants} We recruited 12 participants with various writing experiences from our social network and a mailing list of writers (age range: 20 to 37, M=25; 7 self-identified as male, 4 as female, and 1 preferred not to disclose). On a 5-point Likert scale ranging from 1-never to 5-always, they reported how often they write (Mdn=4), how often they use visuals while writing (Mdn=2), and how often they find visuals to help when writing (Mdn=3). For visual aids, some mentioned using timelines (N=5) and diagrams (N=4). They also reported different writing contexts, including academic (N=9), professional (N=7), and creative (N=6) writing. 

\subsubsection{Apparatus} \label{sec:study1_apparatus}
Participants joined the study remotely via the Zoom conference software and accessed the experiment website using their own computers. The sessions lasted 1 hour and were audio and screen-recorded. Additionally, interactions with the experiment website were also recorded (e.g., features used, commands executed, and answers to questions). Participants received a CAD\$20 gift card for their time. Our institutional ethics board approved the study.

\subsubsection{Tasks} Tasks simulated what writers ask themselves when reviewing their craft. We extracted questions from best practices on revising fiction~\cite{maddenRevisingFictionHandbook1988, nortonDevelopmentalEditingHandbook2011} and then revised them with a professional creative writing instructor. \rev{This led to two open-ended questions per story aspect}: characters (\textit{``Could characters be combined without changing the outcome of the story?''} and \textit{``Is any character too passive?''}); locations (\textit{``Are there any locations which could be removed?''} and \textit{``Are there moments where the spatial logic is broken or unclear?''}); time and rhythm (\textit{``Is there a large gap between two actions that make the story progress?''} and \textit{``Is there a scene you think could take place later/earlier in the story?''}); and focalization (\textit{``If told from another character's perspective, how would the story change?''} and \textit{``What does the main character mention that no other character would?''}).

\subsubsection{Stories} We picked two short stories from the  \textit{Tell Me A Story} dataset~\cite{huotAgentsRoomNarrative2025}, which consists of human-written stories. Stories were screened to be of similar length (\textasciitilde700 words) and featuring more than two speaking characters and location shifts.

\subsubsection{Procedure} After signing the consent form and completing a demographic questionnaire, participants started with one of two conditions: \name or text-only. For the \name condition, they watched a one-minute tutorial and then tried the tool. For both conditions, they read the story fully before starting tasks. \rev{Participants had to think and answer questions aloud within five minutes, and then explain how they came up with the answer}. After each condition, they completed a raw NASA TLX. At the end, they engaged in a semi-structured interview to share impressions, preferences, and strategies. Tasks, stories, and conditions were counterbalanced across participants.

\subsection{Results}
Participants were able to complete the tasks without difficulties in both conditions. However, how they accomplished the task varied. \rev{Themes are based on their relevance to our research questions, as per an interpretivist approach to reflexive thematic analysis~\cite{braunThematicAnalysisPractical2022}.}

\medskip\noindent\textbf{Visualizations helped confirm, analyze, and explore.} All participants mentioned using the visualizations to confirm their intuitions. Visualizations gave the \textit{``reassurance that my hunch is correct, and I'm not missing anything when answering the questions.''} (P6) as they felt \textit{``it was mostly a confidence thing where I felt more sure in what I was saying''} (P2).

The visuals also helped \textit{``have conversations [about] deeper analysis [...] and narrative structures''} (P4). For example, participants used visuals to \textit{``remove this location simply because nobody was there''} (P2), \textit{``identify who can know about each other''} (P5),\textit{``figure out if [an event] was playing a part in the story''} (P6), and see \textit{``how the entities interact with each other''} (P12). P8 added it could be particularly useful \textit{``if you were editing or doing some kind of work on a piece of text where you're like [...] has everyone been given as much time to speak as I want them to.''}

Sometimes, visuals led to discoveries and reflections. For example, P11 mentioned \textit{``it kind of dawned on me by looking at the chronology that [...] we actually don't really get [character]'s motivations''}. Similarly, P6 used the location view to \textit{``reflect on if [location] were to be changed to a different location, would that really have a significant impact on the story? And I realized that's probably no.''} P7 was \textit{``paging through the timeline and seeing how the characters jumped around and kind of weren't even represented [...] it actually showed me [...] that it's like, oh, a lot of moments in this story are sort of disconnected from one another''}. P3 added \textit{``it does make me curious [...] how would a story that features less locations look like''}.

\medskip\noindent\textbf{Visual-driven search vs. skimming.} Participants praised the visuals for their help finding passages. With the visuals, \textit{``it was way more tidier, similar kind of information were in similar places [...] or visually looked similar''} (P5) and \textit{``if I just click [an entity], it'll just highlight the parts where this entity is being talked about [...] I don't have to look at other useless portions''} (P12). In particular, participants contrasted this to their workflow with text, where \textit{``if I just remember, like vaguely, [...] then I can probably pattern-match to the visualization faster than trying to figure out like, what is [...] the keyword I could search to find the sequence of text.''} (P3). In the text condition, participants mentioned how they \textit{``tried to hunt for the dialogues [...] where there were quotations, and tried to see how many of them were said by [character]''} (P4), and were finding locations by \textit{``looking for capitalized nouns [...] so there's always the chance that there was something else [...] like, oh, the coast!''} (P7).

\rev{
\medskip\noindent\textbf{Impact on cognitive load.} Results were inconclusive. Wilcoxon signed-rank did not find a significant difference for mental demand (p=.21), physical demand (p=1), temporal demand (p=.49), performance (p=.38), effort (p=.13), nor frustration (p=.16).

Participants' comments were also mixed. Some mentioned how the visualization reduced their perceived cognitive load because \textit{``[with text] I was trying to visualize the things in my head. But with [visuals], it kind of takes away that mental strain''} (P12) and \textit{``it has all of the characters and locations listed out. If I were to use command F, I would have to put in whatever I can still remember, which oftentimes might not be the complete set''}. Others appreciated being able to focus on other aspects because \textit{``the visualization [...] lets you offload some of the more concrete stuff so that you could explicitly focus on things like character motivations, whereas I found, for the text-based one, I was constantly reskimming in order to refresh this mental schema''} (P11). However, others mentioned that the system added to their cognitive load because \textit{``trying to learn [the tool] [...] almost added this extra layer where I wasn't engaging with the text''} (P7).
}

\medskip\noindent\textbf{Mental model matches and mismatches.} The visuals made sense to some participants because \textit{``I feel like my brain does this, [...] but maybe not this clearly''}. In fact, P9, who had the text condition after the visual condition, mentioned how they \textit{``laid down a timeline in [their] head, something akin to what we've seen in [the visual condition]''} to accomplish the tasks.

Yet, sometimes, there was a clear model mismatch. P8 was the most critical, explaining \textit{``I'm sure there's people who just think like this, and this would be a really useful tool for them, because it is an extension of the way they process information. But for me, it's like really not''}. In fact, P8 mentioned often using visuals when writing but explained their process was different: \textit{``I will write those scenes on sticky notes and then I can start to arrange them in an order that gives me a plot [...], and then I sometimes will use [...] picture boards [...] to remind me of a visual idea or concept''}.

Some other times, the mismatch happened at the semantic level: \textit{``calling it an event for me makes it feel like there's a plot line that's happening. But it wasn't really that.''} (P2) and \textit{``the timeline is very fine grained [...] I wasn't looking at fine-grained scenes [...] but just in terms of more large-scale scenes''} (P3).

\medskip\noindent\textbf{Complementarity of text and visuals} Participants viewed the visualization as complementing the text because \textit{``you still need the text, the words, to fully appreciate the story''} (P1). 
\rev{In fact, when asked how they did the task, participants reported using the views for 74\% of tasks, their memory for 46\%, and reading for 43\% (adds up to more than 100\% because two or more ways could be used).}
One reason is that the visuals pointed to text that then had to be verified. P4 explained \textit{``I was using text for the verification, because there were cases when the visualization was capturing something that could be potentially misleading''}. Another reason is that the visualization could not capture everything. In particular, the visuals only represented explicit information: \textit{``It is really helpful in terms of managing kind of explicit things, like characters, entities, locations [...] But it doesn't quite capture things like motivations''} (P11).

\section{Study 2: Editing and Free-form Writing}\label{sec:study2}
\rev{The vision of \name is to complement the existing writing workflow. As visuals have shown to be helpful for reflecting on the text, we further explore how \name can assist writers during the ``translating'' and revision processes of the cognitive model of writing~\cite{flowerCognitiveProcessTheory1981}}.
\rev{To explore this, we recruited experienced creative writers so that they could give us insights on the possible impact the tool would have on their writing workflows.} 
\rev{This exploratory study used the fully-featured prototype and comprised two parts}: the first part isolates the use of the three different views in a controlled setting, and the second part is a free-form creative writing task. 
We were interested in the following questions: Are the visual representations understood and helpful? Can people express their editing intents using \name? What is the impact of \name on exploration and creativity?

\subsection{Method}

\subsubsection{Participants}
We recruited 8 creative writers from our social network and word-of-mouth (age range: 20 to 31, M=24; 7 self-identified as female and 1 as male). None participated in the first study. All had several years of experience with creative writing, including personal writing (N=7), fiction (N=4), scriptwriting (N=2), and poetry (N=2). On a scale from 1-never to 5-always, they reported how often they use visuals (e.g., mind maps, timelines) (Mdn=2), how often they find visuals to help (Mdn=3), and how frequently they use AI services (Mdn=1) during writing. They reported prior use of visuals to help their writing, including mood boards (N=4), timelines (N=3), mind maps (N=2), and reference imagery (N=2). Participants are labelled W1 to W8 in the rest of the paper.

\subsubsection{Apparatus} Same as \ref{sec:study1_apparatus} except the sessions were about 45 minutes long and participants received a CAD\$15 gift card.

\subsubsection{Tasks} The first part had three task blocks for each view: (1) \textit{entities and actions view tasks} required removing an entity then adding a new entity and action involving the main character; (2) \textit{location and entities view tasks} required moving an entity to an existing location, creating a new location, and move another entity to it; (3) \textit{timeline view tasks} required reordering events. Instructions were phrased as questions, forcing participants to engage with the story. For example, a timeline task asked \textit{``What if Jack loses his hat when he flies into the sky?''}. This required participants to understand the story's events and move the action to the right location. Tasks were counterbalanced across participants. The second part involved a single writing task using a writing prompt. 

\subsubsection{Procedure} After completing the consent form and demographics questionnaire, participants watched a 30-second video introducing the tool. Before a task block, they watched a short video demonstrating the relevant features. 
After a block, participants rated their perceived success on a 5-point semantic differential scale and answered five questions related to usability, understandability, and workflow integration. Following all tasks, participants engaged in a free-form exploration to continue a story with no instructions other than to ``explore what could happen next.'' At the end, participants completed an exit questionnaire (including the creativity support index (CSI~\cite{cherryQuantifyingCreativitySupport2014}) and the raw NASA TLX) and engaged in a semi-structured interview to gather their experience and perceived advantages and disadvantages of the system.

\subsection{Results}
Participants completed the tasks and the free-form exploration without major difficulties. Below, we summarize the quantitative findings using descriptive statistics. Following recommendations for similar study designs~\cite{zhuAssessingComparingAccuracy2018, massonStatslatorInteractiveTranslation2023}, confidence intervals use the studentized bootstrap with 10,000 repetitions. \rev{Themes are based on their relevance to our research questions, as per an interpretivist approach to reflexive thematic analysis~\cite{braunThematicAnalysisPractical2022}.}

\begin{figure*}[h]
    \includegraphics[width=\textwidth]{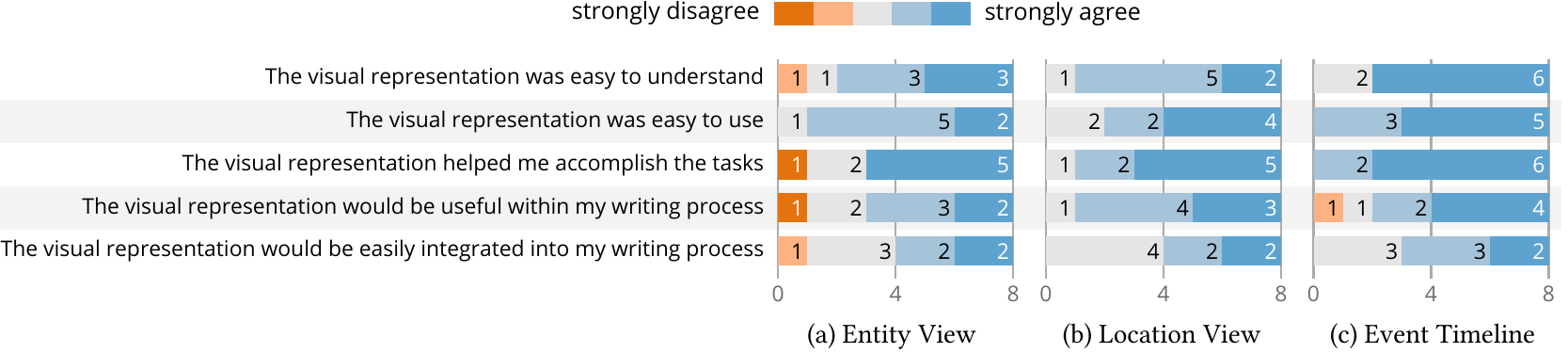}
    \caption{Participants' response for study 2 when rating the 5-point scale statements for (a) the entities and actions view; (b) the locations and entities view; and (c) the event timeline.}
    \label{fig:statements}
\end{figure*}

\begin{table}[]
\begin{tabular}{lcc}
\hline
Scale               & Avg. Score (SD) & Avg. Counts (SD) \\ \hline
Enjoyment            & \textbf{15.73} (1.84)          & \textbf{3.50} (1.00)      \\
Immersion            & 12.18 (3.04)          & \underline{3.13} (1.76)      \\
Results Worth Effort & \underline{15.28} (2.26)          & 3.00 (1.50)      \\
Exploration          & 13.48 (2.91)          & 2.88 (1.27)      \\
Expressiveness       & 13.70 (2.86)          & 2.50 (0.50)      \\  \hline
Overall CSI Score    & 71.50 (21.12)         &                  \\ \hline \\
\end{tabular}
\caption{Results for the Creativity Support Index (CSI) sorted based on the average factor count (i.e., the factors considered most important by participants). The highest value is in bold, and the second highest is underlined. Like previous work~\cite{sangho_luminate_2024,carroll_creativity_2009}, we omit the \textit{Collaboration} factor to avoid confusion.}
\label{tab:csi-result}
\end{table}

\medskip\noindent\textbf{Finding scenes and keeping track of events, entities, and locations.} As a search tool, the results of this study corroborated those of the first study. Most participants agreed that ``The visual representation was easy to understand'' for all three views (\Cref{fig:statements}). W5 appreciated that you can \textit{``click on a person, click on a location, and then see stuff pop up''} instead of \textit{``having to look through [...] many paragraphs [...] or doing keyword search''}. W4 explained \textit{``Just having a visual representation helps, you know, just with the different perspectives''}. W3 added \textit{``Moving around the timeline was a good way of, kind of figuring out different events''}. W1 and W5 also mentioned how they usually struggle to keep track of everything while writing their stories and that the system could help them. For instance, W5 details that currently they \textit{``have booklets everywhere, and [they] have it annotated weird [...] it's a mess and this is so much easier. You got like, you know, your locations, your people [...] keeping track of things. Organizing who's where, what's going on, being able to scroll through, like highlighting part of the text shown on the timeline [...] making it easy to understand''}. \rev{This was also reflected in participants reporting low mental demand on the NASA-TLX (M=7.49 SD=1.95).}

\medskip\noindent\textbf{Specifying temporal, spatial, and entity-related edits.} 
On a 5-point semantic differential scale from 1-unsuccessful to 5-successful, participants rated their success M=4.25 (95\% CI: [3.8, 4.6]) at accomplishing the tasks. They were most successful at the task about adding and removing entities (M=4.5, 95\% CI: [3.8, 5]), changing the locations of entities (M=4.25, 95\% CI: [3.4, 4.8]), and reordering events (M=4, 95\% CI: [2.8, 4.7]).
\rev{Participants used an average of 2.42 visual edits to accomplish a task. In the free-form part, the most common visual edits were adding actions (36\%), editing actions (26\%), and editing traits (19\%).}
Specifically, participants appreciated that the edits produced text as a good starting point. W2 mentioned \textit{``If I add in an extra character, it gives you just enough text to build that story with that new character''}. W4 appreciated that it fixed inconsistencies when editing locations: \textit{``the fact that it also updated the sentence where the hat flew off his head that is extremely useful because that’s the sort of inconsistencies that can easily creep in when you’re making modifications like that''}. Similarly, W6 expressed, \textit{``I really liked the ability to create relations between characters''.}

\rev{On the NASA-TLX, they also reported high-performance (M=7.49 SD=1.95) for low effort (M=3.39, SD=2.17), frustration (M=2.5, SD=1.2), temporal demand (M=1.18, SD=1.04), and physical demand (M=1.18, SD=1.04).}
For example, W6 further expressed that \textit{``[the tool] could be useful as a touch-up tool for when I want to make larger sweeping changes to a very long existing story [...] I can confirm whether or not I want to keep each of these edits''} stating that the system \textit{``made it a lot easier than any sort of other systems that I’ve used''}. The idea of reviewing the changes was important for participants, as W3 explained \textit{``You’d still have to review it, but it made enough changes that it [the story] was able to flow well.''}

\medskip\noindent\textbf{Helping explore and be creative.} 
Participants generated an average CSI score of 71.5 (SD=21.12) (\Cref{tab:csi-result}).
They felt the tool was useful as a creative assistant, even when the text produced was not up to their standards.  W6 explained \textit{``A lot of times, I am interested in how making a change to a given detail in a story would look [...] I was appreciating [the tool] as sort of an exploratory tool where even if it doesn’t end up producing something that I would personally publish, I think it’s really good to give an example that I can then engage with''}. Similarly, W5 mentioned \textit{``You can just like, throw in a verb or the general action, interaction between them and it was kind of spitballing ideas''}. Regarding the timeline, W7 felt that \textit{``it’s an easy way to explore your options of what it sounds like if something happens here or if it sounds better when it’s there.''}. They also mentioned how just looking at the visualization could help, for instance: \textit{``In terms of ideas, I could see how looking at the map, something could pop up into your head […] once you have, like, those objects laid out on a map, then that can help you spark inspiration''}.

\medskip\noindent\textbf{Impact on the writing workflow.} Perhaps due to the study setting, all participants heavily used the system during the free-form part. Most participants used the system to explore, \textit{``just to see what would happen''} (W2), then read the suggested changes and kept them or undid them either by editing an event (W2, W6), undoing using the history graph (W3), deleting actions and characters (W1, W4, W6), or rewriting passages manually (W4, W8). When exploring, participants mostly used the actions and the location views. The timeline was only used to select specific scenes to edit. Participants also had different strategies when creating events, for example, W1, W4, and W7 wrote longer and detailed actions such as \textit{``Interrupts conversations, she says that she saw it too, and she is also confused''} (W1) or explicitly writing a dialogue \textit{``Says `what are you doing here?'''} (W7), whereas others wrote one or two-word actions or descriptions. W8 had the most unique workflow in that they used the visuals to edit the text only once. Instead, they spent most of the session writing their story manually and would periodically refresh the visualization based on what they wrote and then reflect. They explained \textit{``In my most focused state I would just be focused on writing, but then when I was done, I was out of it, I would scan back and forth between text and the [visuals]''}.

\medskip\noindent\textbf{Mismatches between the system and participants' existing workflows.} Questions related to integrating the system in the writing process were rated the lowest (but still with an Mdn=4). Similar to the findings from the first study, the most critical participants explained that the visuals did not match their workflow and the visualizations they already used. W4 mentioned \textit{``I really really like having a visual representation of the story, but this is not the visual representation I would choose [...] I would like a combination of those two things [timeline and entity view] where those interlinks and connections appear within the timeline.''}\footnote{While this visualization was not implemented in our system, it can be modelled with our framework: start with characters, connect by events, position by locations, and unfold by time}. Similarly, W6 explained how a single timeline would not be enough for them: \textit{``my brain interprets timelines in stories where I enjoy having things happening in multiple places at the same time and having a single linear approach makes it very hard to have that sort of interesting kind of development''}\footnote{Again, this can be modelled by our framework by unfolding by both time and temporality to visualize nonlinear narratives}. Other than visual representations, W4 also mentioned they \textit{``don't want the system overwriting [...] because then it's not me expressing myself through text''} and instead, they would prefer \textit{``just highlighting the places where things need to change, or the things are inconsistent''}. 

\medskip\noindent\textbf{Participants would have liked more control in using the visual representations to edit the story.} While participants could express their intents with the tool, there are nuances in how an idea can be expressed in terms of writing style and voice, which is not currently supported by our system. For example, when discussing adding actions by connecting two entities, W8 mentioned \textit{``I feel like that’s very limiting how you wanted to express what you want versus just typing it out in a sentence because it kind of format it in a specific manner''}. Similarly, W7 explained \textit{``in terms of expressing myself, I would say that it does help [...] but I don’t think it allows for like total freedom of expression [...] if you had something, in your mind of how you wanted to say it, it might not come out that way''}.

\section{Discussion}\label{sec:discussion}
We propose \name, an approach that uses visual representations of story elements to support writing and revising narrative texts. Despite how little writing support tools have used visualizations~\cite{zhao2025making}, our findings show that they hold great promise to help review and ideate stories. The space of possibilities is large, and our framework of story constructs is an initial attempt at structuring it. In our study, we looked at a subset of this space, using simple representations that did not cover all story elements. However, we encourage others to continue this line of research.

One question with tools like this is how it might change the role of writers.
AI-assisted writing tools tend to make us write less and review and edit more~\cite{buschekCollageNewWriting2024a}. 
We believe \name can accommodate both traditional and ``editorial'' writing roles: in its simplest form, \name is automatizing the creation of visualizations writers were \textit{already} using~\cite{zhao2025making, ackermanEfficientWriterUsing2017, neuwirthRoleExternalRepresentation1989, SaveTheCat, Articy}, with the added bonus of serving as input, if the writer desires it.

Another question is how \name could be integrated with other writing support tools. We view \name as complementary to tools that rely on textual prompts (e.g., ChatGPT or Grammarly). Like work on direct manipulation with LLMs has shown~\cite{massonDirectGPTDirectManipulation2024}, we expect \name to be preferable when writers have a clearer idea of the edit they want to perform (i.e., ``move this character there'') because it helps refer to story elements unambiguously. Whereas textual prompts are preferable when the goal is vague (e.g., ``make this story sound better'').

Finally, our work targets the creative writing of narratives that involve tangible elements like characters and locations. It applies not only to novels but also to screenwriting for films, TV shows, video games, and theatre plays. However, it is unclear how \name could work in domains like academic writing, where the elements manipulated are not so tangible.

\subsection{Future Work}

\subsubsection{Extending the framework and exploring new visual interactions}
Our framework (\S\ref{sec:taxonomy}) operates at a different conceptual level than typical visualization frameworks, such as the grammar of graphics~\cite{wilkinsonGrammarGraphics2005}. It combines story elements into constructs that are then linked to visuals without prescribing how to extract story elements or what the visualizations should look like, allowing more freedom in implementation. For example, our system represents locations as bubbles instead of a traditional map. Alternatively, images could indicate what an object, character, or location looks like, allowing manipulation (e.g., changing a character's hair colour) or emphasis on certain descriptive points. More work is needed to identify which constructs can best serve writers within \name workflows, and how to best represent them. \rev{
For example, inter-character relationships are important in a story~\cite{bochnerRelationshipsStories1997} and a view constructed by connecting actors would help visualize and edit these relationships.} Additionally, the framework can be extended to include story elements beyond those highlighted by Bal~\cite{balNarratologyIntroductionTheory1997}. 
For example, "emotions" could be a basic building block, as used in StoryPrint~\cite{watsonStoryPrintInteractiveVisualization2019}, and "motivations" could be another, as suggested by P11 during the first study. 
\rev{In fact, other narratology models, such as Propp's morphology of the folk tale~\cite{proppMorphologyFolkTale2010}, Todorov's principles of narratives~\cite{todorov2PrinciplesNarrative1971}, or the contemporary concepts on narrative time by Hume~\cite{humeFantasyMimesisResponses2014} would lead to a different framework. }

\subsubsection{Visual editing of writing style and story plots} Creative writing encompasses more than just events, actors, locations, and time. The approach could also help edit other aspects, such as writing style. As proposed by \citet{kempenAuthorEnvironmentsFifth1987}, manipulating the syntax tree can restructure sentences. Alternatively, tone, sentence length, or structure could be represented with histograms or text embellishments, allowing style adjustments through visual manipulation.
At a higher level, \name can aid revising the plot. Similar to work on manipulating outlines~\cite{dangTextGenerationSupporting2022, zhangVISARHumanAIArgumentative2023}, main plot events can be visually represented to be reordered or changed.

\subsubsection{Suggestions and creativity support in the visual space} 
The mental processes involved in interpreting visuals differ from those involved in processing language~\cite{paivioMentalRepresentationsDual1990a}. Future work could explore the impact on the creativity of writing stories through visuals rather than with text. Similarly, typical writing support features would be interesting to adapt to the visual space. For instance, there are many benefits of phrase suggestions~\cite{buschekImpactMultipleParallel2021, bhatInteractingNextPhraseSuggestions2023}, and future work could investigate if there are similar benefits to suggestions in the visual space. These visual suggestions would essentially try to ``auto-complete'' an interaction initiated on a visual element.

\rev{
\subsubsection{Supporting long stories} For long stories, the visuals have to find the right balance between overview and details. Our prototype attempts to show the details of all the events and entities and rely on interactions to declutter the views, by merging overlapping actions and allowing users to hover or select a passage to filter what is shown. If details are prioritized, this view could be unfolded by temporality to “spread out” the interactions over time. Unfolding again by time would visualize complex nonlinear narratives. Alternatively, interaction techniques such as details-on-hover or semantic zooming could also help manage longer stories.
}

\rev{
\subsubsection{Visualization-builder to support custom visuals} The studies showed creative writers have various preferences in terms of visuals. One way to support these workflows would be to help users construct their own visuals. Our procedural framework could inform such a view-builder: the eight story elements would be the basic building blocks, and users could construct a view by applying the different operators. Once the view is constructed, it could become interactive and automatically populated from the text.
}

\subsection{Limitations}
\rev{
\subsubsection{Participants' experience and reservation towards AI might have impacted our results} In study 1, half the participants had no creative writing experience, meaning they were possibly at a disadvantage to accomplish the tasks. Conversely, all participants in study 2 had years of creative experience, and, while they could comment on how \name could be integrated in their workflows, the results could not tell us about the benefits of the tool for beginners.
Similarly, some participants expressed reservations towards AI and reported rarely, if ever, using it. W4 explained \textit{``I write because I want to, so if I want to be writing then there’s ideas that I have, and I want to express them. I don’t want a system to tell me what it is that I am going to be expressing''}. Considering writers' have strong opinions about AI~\cite{hertzmannCanComputersCreate2018,biermannToolCompanionStorywriters2022}, this might have biased the results.
}

\subsubsection{The system might not have been robust enough to test the full potential of \name} The system relied on an LLM that occasionally responded unexpectedly, such as modifying unintended elements or refusing to do certain modifications. Additionally, some manipulations were slow to execute (up to 10-15 seconds) despite our efforts to parallelize requests. These issues may have affected our results as interactions should be seamless and paired with immediate effects to promote user exploration~\cite{massonSuperchargingTrialErrorLearning2022, shneidermanCreativitySupportTools2006, shneidermanDirectManipulationStep1983}.

\section{Conclusion}
We proposed \name, an approach to support writing by reviewing and manipulating visual representations. In doing so, we defined a framework to help inform the design of story visualizations.
By applying this framework, we implemented a prototype system demonstrating one possible design for \name tools. 
Two studies covering the different aspects of the writing process showed the potential of creative writing to help keep track of story elements, rapidly specify edits, and explore story variations in a way that encourages creativity. 
Broadly, our work advocates for a new generation of writing support tools that embed visualization to help review and edit text narratives.


\begin{acks}
We thank Daniel Aureliano Newman for his help revising the questions from study 1.
This project was undertaken thanks to funding from IVADO, the Canada First Research Excellence Fund, and NSERC Discovery Grant RGPIN-2018-05072.
\end{acks}

\bibliographystyle{ACM-Reference-Format}
\bibliography{_references.bib, zotero_do_not_modify}


\appendix
\section{Appendix}

\subsection{Extracting Information from the Text}\label{appendix:extraction}
The visual representations are generated from information extracted from the text about the entities, locations, and events. The extraction follows three steps, in sequence: 1) extract the entities and their traits; 2) extract the locations; 3) extract the events for each sentence. For all steps, OpenAI's ``GPT4-o'' model was used.

\subsubsection{Extracting Entities}
The prompt below extracts all the entities (i.e., characters and inanimate objects) from a story.
\begin{prompt}
<STORY TEXT>\\
Extract all the entities in this story.
For each entity, extract its `name', an emoji best visually
describing the entity (e.g., use the emoji of a person if it 
is a person but avoid reusing the same emojis),
and properties about the entity, if any (no more than 3).
Properties have to be adjectives describing the entity and
their value should represent the intensity of the property 
(on a scale from 1 to 10).
\end{prompt}

The model responds with a structured output using the following JSON schema: 
\begin{json}
entities: [{
    name: string,
    emoji: string
    properties: [{
        name: string,
        value: number
    }]
}]
\end{json}

\subsubsection{Extracting Locations}
The prompt below extracts all the locations from a story.
\begin{prompt}
<STORY TEXT>

Extract all the main locations visited by the characters in this story.
For each location, extract its `name' and an emoji best visually representing the location
\end{prompt}

The output is structured, using the following JSON schema:
\begin{json}
locations: [{
    name: string,
    emoji: string
}]
\end{json}

\subsubsection{Extracting Events}
Extracting the events is done after the entities and locations are extracted. First, the story is divided into sentences. Then, in parallel and for each sentence, the following prompt is used.
\begin{prompt}
BEFORE: <TEXT BEFORE> \\
TEXT: <SENTENCE>\\

\medskip Extract the actions done by the characters in TEXT and only the actions in TEXT. Do not extract the actions from BEFORE.
Only consider actions that are happening exactly at the moment of TEXT, ignore memories etc.
If there are no actions fulfilling these criteria in TEXT, then return an empty array.
Source and target should be characters from this list: <ENTITIES>.
Here are some possible locations but there might be others: <LOCATIONS>
If an action is done by a character to itself, then the source and target character should be the same.  For each action, extract the `name' of the action (no more than 2 words),
the source character (the character doing the action) and the target character (the character targeted by the action), and the location of the action (you can use `unknown' if the location cannot be inferred from the text).
\end{prompt}
With <TEXT BEFORE> corresponding to the text before the sentence (to give the model some context), <SENTENCE> the text of the sentence, <ENTITIES> the list of entities extracted previously, and <LOCATIONS> the list of locations extracted previously.

The output is structured, using the following JSON schema:
\begin{json}
actions: [{
    name: string,
    source: string,
    target: string,
    location: string
}]
\end{json}

All extracted events are associated with their respective sentences to support highlighting the text when manipulating the visual representations. Additionally, to make future extraction faster, only sentences that changed are re-extracted.

\subsection{Editing the Story by Manipulating the Visuals}
Below, we list the prompts used to suggest edits to the story when visual representations are modified.

\subsubsection{Reorder Events in the Timeline}
This prompt works by giving the list of events in the current and new order. Once the story is modified, the newly generated text is re-extracted fully to make sure the visual representations match the new text.
\begin{prompt}
<STORY TEXT>\\
In this story, the current order of actions is:\\
<CURRENT ORDER>\\
Rewrite the story to EXACTLY follow this new order:\\
<NEW ORDER>
\end{prompt}
<STORY> is the full text of the story, <CURRENT ORDER> is a list of the events in the story, as extracted previously, and <NEW ORDER> is this same list but with the events reorganized. Events are represented in the form ``<SOURCE> <EVENT NAME> <TARGET>''.

\subsubsection{Adding, Changing or Removing An Action} All the prompts resulting from manipulating links between nodes on the event view follow the same format:
\begin{prompt}
<STORY TEXT>\\
SOURCE: <SOURCE ENTITY>\\
TARGET: <TARGET ENTITY>\\
ACTION: <ACTION NAME>\\
<PROMPT>
\end{prompt}

<PROMPT> is either ``Rewrite <blank> to add that SOURCE also <ACTION> TARGET'' when adding an action, ``Rewrite the story so that SOURCE also <ACTION> TARGET'' when changing an action, and ``Rewrite the story so that SOURCE does not do ACTION to TARGET'' when removing an action.

\subsubsection{Removing an Entity}\par
\begin{prompt}
<STORY TEXT>\\
Rewrite the story so that there is no <ENTITY NAME>
\end{prompt}

Note that no prompts are executed when adding an entity. This is only when adding an action that the text is modified.

\subsubsection{Moving an Entity}
\begin{prompt}
<STORY TEXT>\\
Rewrite the story so that <ENTITY NAME> never goes to the <CURRENT LOCATION> but instead goes to the <NEW LOCATION>
\end{prompt}

\subsubsection{Targeted Edits}
Visual edits can modify only a specific passage or set of scenes, if those scenes are selected on the timeline. For this to work, the prompts mentioned above are slightly modified so that <STORY TEXT> includes the fully story text with the selected passage replaced by ``TEXT\_TO\_REWRITE''. Then, a new line is added defining ``TEXT\_TO\_REWRITE'' as the selected text passage. The prompt is then modified to indicate that it should only rewrite the passage. The result is then integrated into the full story. This approach is inspired by the prompts used in DirectGPT~\cite{massonDirectGPTDirectManipulation2024}.

\end{document}